\documentclass[aps,prb,twocolumn,superscriptaddress,showpacs]{revtex4}

\usepackage{graphicx}
\usepackage{amsmath}
\usepackage[applemac]{inputenc}

\begin{document}

\title{Resonant quantum tunneling of spin chains in a three-dimensionnal magnetically ordered state}

\author{E. Lhotel}
\affiliation{Centre de Recherche sur les Tr\`es Basses Temp\'eratures, CNRS, BP~166, 38042 Grenoble, France}
\email[Corresponding author: ]{elhotel@cea.fr}
\author{E. N. Khatsko}
\affiliation{Institute for Low Temperature Physics and Engeneering, National Academy of Sciences of Ukraine,  310 164 Kharkov, Ukraine.}
\author{C. Paulsen}
\affiliation{Centre de Recherche sur les Tr\`es Basses Temp\'eratures, CNRS, BP~166, 38042 Grenoble, France}

\begin{abstract}
We show that resonant quantum tunneling of the magnetization, until now observed only in zero-dimensional (0D) cluster systems, occurs in the molecular Ising spin chain $[(CH_3)_3NH]CoCl_3\cdot 2H_2O$, which orders as a canted 3D-antiferromagnet at $T_C=4.15\ K$. We present three features that prove unambigously the existence of resonant quantum tunneling below $0.3\ K$ ($T_C/12$): Temperature independent relaxation of the magnetization, speeding up of the relaxation at a well defined field ($1025\ Oe$) and increase of the magnetization each time this field is crossed during a succession of minor hysteresis loops. A mechanism is proposed to describe the behavior based on  a simple model of 0D domain wall nucleation.
\end{abstract}


\maketitle

The existence of resonant quantum tunneling of the magnetization (QTM) for clusters of several strongly correlated spins is well established. The archetypes of these zero-dimensional (0D) systems, called single molecule magnets (SMMs), are $Mn_{12}$ \cite{Sessoli93} or $Fe_8$ \cite{Sangregorio97}. Each cluster carries a well defined macrospin $S=10$ with a large Ising like anisotropy and is weakly coupled to its neighbors. QTM is observed at low temperatures when the system displays a crossover from a thermally  activated relaxation over the anisotropy barrier to a temperature independent relaxation due to tunneling through the barrier. In addition, a resonance effect occurs as a function of field at level crossings of the $2S+1$ multiplet. 
Recently, studies of the effect of weak dipolar and exchange interactions on tunneling have been reported. 
When considering only two clusters in the case of the $[Mn_4]_2$ dimers, the weak exchange between $Mn_4$ units has been found only to modify, not suppress QTM and resonance features \cite{Wernsdorfer02}. A similar behavior is observed for $Mn_4$ clusters weakly coupled along a chain in the presence of a partial ordering \cite{Wernsdorfer05a}. But for 3D $Mn_4$ networks, no resonance effects were reported in the superparamagnetic blocking regime even though (incoherent) magnetic quantum tunneling could be inferred from low temperature specific heat measurements  \cite{Evangelisti04}. For 1D systems, the so called single chain magnet $Mn_2Ni$ shows dynamics with a quantum regime below $700\ mK$ but an absence of resonant effects \cite{Wernsdorfer05}. 

In the present letter we will show that the well known compound $[(CH_3)_3NH]CoCl_3\cdot 2H_2O$, (CoTAC), a 3D network of exchanged-coupled ferromagnetic chains, does exhibit QTM  \textit{and} resonance effects  \textit{well below} its 3D antiferromagnetic transition temperature ($T_C=4.15\ K$). Note that there is an essential difference between this system and those previously studied since CoTAC undergoes a classic  \textit{3D phase transition}.
We use standard experimental procedures which have proved their effectiveness in the search for QTM and resonance effects for SMMs. We measured the absolute values of the magnetization of a single crystal of 12.2 mg by the extraction method, using a superconducting quantum interference device magnetometer equipped with a miniature dilution refrigerator developed at the CRTBT, CNRS. Special care was taken to check the accuracy of the alignment of the crystallographic axis ($a$ or $c$) of our sample along the field axis, which is necessary due to its large anisotropy. 

\begin{figure}
\includegraphics[width=5cm, keepaspectratio=true]{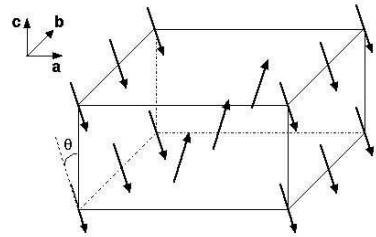}
\caption{Representation of the magnetic structure of CoTAC. The canting angle $\theta$ is equal to $10^{\circ}$. }
\label{fig1}
\end{figure}

We recall that CoTAC crystallizes in the $Pnma$ group \cite{Losee73}. Neutron diffraction \cite{Bruckel93}, electronic paramagnetic resonance \cite{Kobets02} and early susceptibility \cite{Groenendijk82} results provide a complete description of the magnetic structure (see Fig.~\ref{fig1}). The ferromagnetic chains run along the $b$ axis. The spin $S=3/2$ of the  $Co^{2+}$  ion is usually described as an effective spin $S_{eff}=1/2$ (with  $g=7.5$) and is perpendicular to the $b$ axes and tilted from the $c$ direction by an angle $\theta=±10^{\circ}$. The $c$ axis is the easy direction of magnetization with axial and transverse anisotropy equal to $D=1.88\ K$ and $E=0.422\ K$ respectively.
The ferromagnetic exchange interaction energy $J_b=14\ K$ along $b$ is much larger than the inter-chain couplings: ferromagnetic $J_c=0.14\ K$ and  antiferromagnetic $J_a=-0.008\ K$. The system behaves as a collection of independent ferromagnetic Ising chains at high temperatures. However, $J_c$ and $J_a$ are large enough to cause  three-dimensional magnetic order below the critical temperature $T_C$ of $4.15\ K$. The low temperature state is a canted antiferromagnet exhibiting a net ferromagnetic component along the $a$ direction. 

The weak ferromagnetism that develops along the $a$ axis motivated previous magnetic studies performed above $1.4\ K$. In particular, ac susceptibility measurements (ac field parallel to $a$) have shown different regimes for the dynamics depending on the temperature with a slow relaxation below $T_C$, which was attributed to domain-wall motion \cite{Okuda81}. Our measurements along the $a$ axis are consistent with these results.

\begin{figure}
\includegraphics[width=8.5cm, keepaspectratio=true]{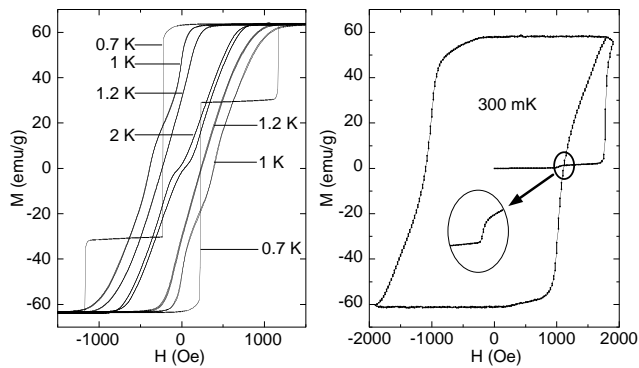}
\caption{Left: Major hysteresis loops in the $c$ direction at a field sweeping rate $dH/dt=76\ Oe.s^{-1}$ at four temperatures. 
Right: Hysteresis loops measured at $300\ mK$ with field increments of $25\ Oe$ every 34 s ($\sim dH/dt= 0.74\ Oe.s^{-1}$). A  small step in the first magnetization curve is observed (the vertical axis is magnified by a factor of 10 in the circle). A large step is seen in the major hysteresis loop at the same field: $dM/dH$ \textit{vs.} $H$ presents sharp peaks for both curves at the same field $\approx 1025\ Oe$ where the magnetization steps occur (not shown). }
\label{fig2}
\end{figure}

In order to observe possible QTM effects we carried out an extensive study at lower temperatures. The focus of our investigation was with the applied field and magnetization measurements along the $c$ axis, which we will discuss for the remainder of the paper. Shown in the left side of Fig.~\ref{fig2} are plots of major hysteresis loops measured with a field sweeping rate of $76\ Oe.s^{-1}$ and  taken at different constant temperatures ($0.7 \leq T \leq 2\ K$). Hysteresis loops occur for this field sweeping rate below approximately $2.5\ K$, and become wider as the temperature decreases. Below $1\ K$, two reproducible avalanches of the magnetization (that is to say, a fast reversal of the magnetization at a given field) appear for each branch of the loop together with a sudden and large increase of the sample temperature: See, for example, $M(H)$ at $0.7\ K$ in Fig.~\ref{fig2} (left). However, by ramping the field at a rate slower than approximately $2\ Oe.s^{-1}$, the avalanches disappear altogether. The slower field ramping allows the heat that is released by certain fast relaxation process to be dissipated in the sample and adsorbed by the mixing chamber without excessive local heating, which we believe is most likely the cause of  the avalanches. Although the avalanches are interesting in their own right,  below we will  adopt this slower field ramping rate in order to study the dynamics at low temperature. 

\begin{figure}
\includegraphics[width=6.5cm, keepaspectratio=true]{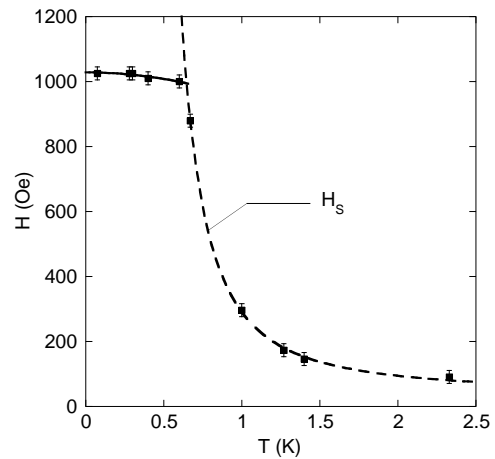}
\caption{$H_s$ \textit{vs.} $T$ from zero-field-cooled curves, for $dH/dt=1\ Oe.s^{-1}$. The dashed line is a fit to the equation : $H_s=30.6 \exp(E' /k_B T)$ with $E'/k_B=2.2\ K$. }
\label{fig3}
\end{figure}

At $2\ K$, an inflection point at approximately $70\ Oe$ can be seen superimposed on top of the hysteresis loops. This is the signature of the metamagnetic transition previously reported from magnetization \cite{Groenendijk82} and NMR \cite{Spence74} studies performed above $1.4\ K$. 
Instead, at very low temperature, a step in the major hysteresis loops is observed [See Fig.~\ref{fig2} (right)]. This step occurs at a field of $1025\ Oe$, thus well above the metamagnetic transition field at $2\ K$. Using the standard procedure by starting from the Zero Field Cooled (ZFC) state $M=0$ and measuring the first magnetization curve $M_{ZFC}(H)$, the same feature is observed (See Fig.~\ref{fig2} right). We explored the temperature dependence of the onset field $H_{s}$ of the step (or the inflection point, according to the temperature) appearing in the $M_{ZFC}(H)$ curves.  From $3.5$ to $1.2\ K$,  $H_{s}$ varies little with temperature. Between $1.2$ to $0.7\ K$, there is an abrupt increase  to approximately $1000\ Oe$ (see Fig.~\ref{fig3}). In this temperature range, $H_{s}$ increases by a factor of 20, obeying roughly an exponential function of $1/T$: $H_{s}\approx 30.6\ Oe \times \exp(E'/k_B T)$ with $E'/k_B=2.2\ K$. More interestingly, the rapid increase in $H_{s}$ stops suddenly below $0.7\ K$ and tends to reach a constant value of approximately $1025\ Oe$, below $0.4\ K$. Note that this low temperature field value ($1025\ Oe$) is the same for the first magnetization curves  as well as for major hysteresis loops [See Fig.~\ref{fig2} (right)]. 

\begin{figure}
\includegraphics[width=7.3cm, keepaspectratio=true]{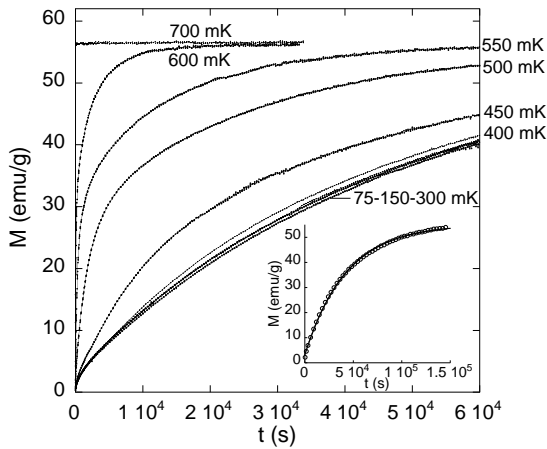}
\caption{$M$ \textit{vs.} $t$. Relaxation curves recorded over almost 17 h at different temperatures between $75$ and $700\ mK$ in a $1025\ Oe$ dc field. The initial state is a ZFC state. Inset: Relaxation curve at $75\ mK$ at long times ($t<42$ h). The line is an exponential fit.}
\label{fig4}
\end{figure}

\begin{figure}
\includegraphics[width=6.5cm, keepaspectratio=true]{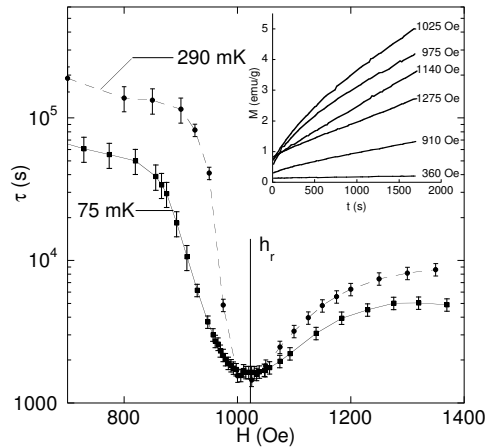}
\caption{Effective relaxation time $\tau_{eff}$ as a function of field $H$ at $75$ and $290\ mK$ in a semi-logarithmic scale deduced from the ($t<30\ min$) relaxation curves $M(t)$ such as those shown in the inset for several fields at $75\ mK$.  }
\label{fig5}
\end{figure}

We made a systematic investigation of this unusual behavior by measuring the relaxation of the ZFC magnetization as a function of temperature at the step field (see Fig.~\ref{fig4}) as well as a function of field near the step at constant temperature (see Fig.~\ref{fig5}). Each curve was obtained by first zero-field-cooling the sample; then a measuring field was applied and the magnetization $M(t)$ was recorded as a function of time while the magnetization relaxed (\textit{i.e.}, upward in the figures). Figure~\ref{fig4} shows typical relaxation curves of $M(t)$ measured at $1025\ Oe$ for several temperatures, from $700$ down to $75\ mK$. Above $700\ mK$, the relaxation is very fast. As the temperature decreases, the relaxation becomes increasingly slower which is common  behavior for thermally activated processes. However, below $300\ mK$ the relaxation becomes independent of temperature. In fact, between $75$ and $300\ mK$, all of our relaxation curves collapse onto one. 

The inset of Fig.~\ref{fig5} shows the field dependence of the relaxation of the magnetization at $T=75\ mK$.  For field values at the step field $1025\ Oe$, the data can be reasonably well fitted with a simple exponential, but moving away from this field, the fits become quite poor. Nevertheless we can extract an approximate relaxation time $\tau_{eff}$  from the fits which are plotted in Fig.~\ref{fig5}  as a function of applied field for  $T=75$ and $290\ mK$. Curiously, far from $1025\ Oe$, the relaxation for a given field at $T=75\ mK$ is faster than at $290\ mK$. However, the important feature is that the relaxation times  show a deep minimum or resonance  at $1025\ Oe$ (note that the vertical scale is logarithmic). 

These results are very similar to resonant QTM observed in SMMs such as $Fe_8$ \cite{Sangregorio97}. However, for CoTAC, there is no resonance at $H=0$; the first occurs at  $1025\ Oe$, and with the present experimental setup, we could not confirm the presence of other resonances at higher fields due to the fast relaxation and problems with avalanches. Note that the first resonance is not symmetrical, and is about $200\ Oe$ in width, thus comparable with that observed for  $Fe_8$ \cite{Ohm98}. In the latter case, the resonance width has been explained in terms of hyperfine fields inside the molecule and dipolar fields acting from one giant spin to another. 

\begin{figure}
\includegraphics[width=6.5cm, keepaspectratio=true]{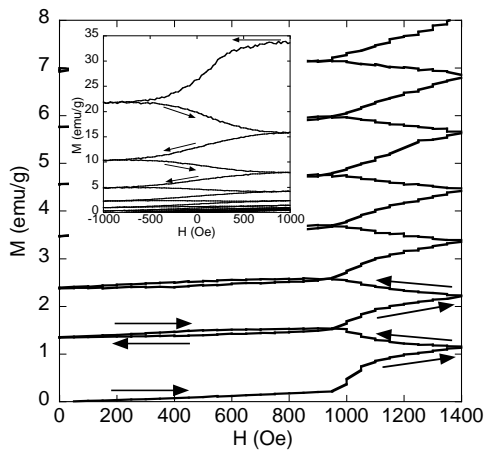}
\caption{$M$ \textit{vs.} $H$. After ZFC, minor hysteresis loops at $300\ mK$ between $0$ and $1400\ Oe$, with a field step of $50\ Oe$ every 34 s ($\sim dH/dt= 1.5\ Oe.s^{-1}$). More relaxation in the magnetization is observed each time the field crosses $1025\ Oe$. Inset: minor hysteresis loops around $H=0$ (first resonance) for the SMM $Fe_8$. }
\label{fig6}
\end{figure}

Another comparison between CoTAC and $Fe_8$ can be found from a study of the effect of minor hysteresis loops around the presumably resonant field $h_r=1025\ Oe$. Figure~\ref{fig6} shows loops measured at $300\ mK$, starting from the ZFC state. The field was increased in  increments of $50\ Oe$ every 34 s ($\sim dH/dt= 1.5\ Oe.s^{-1}$). The magnetization is nearly flat starting from H=0 but increases sharply in the vicinity  of the step, and slows down again for larger fields. On decreasing the field, the magnetization increases again near the step, and slows down dramatically on the other side, and so on. Note that the behavior is not symmetric about the step field, which is consistent with the variation of $\tau_{eff}$ shown in Fig.~\ref{fig5}.
Nevertheless, the resemblance to $Fe_8$ can be seen in the inset of Fig.~\ref{fig6} where minor hysteresis loops  are shown for a single crystal of $Fe_8$. These data were taken by first saturating the sample, and measuring about the first resonance at $H=0$. 

In summary, using the same experimental procedures,  we have obtained results analogous to those previously observed in the SMM $Fe_8$ which are characteristic of resonant QTM at low temperature:  temperature independent relaxation of the magnetization below $T=300\ mK$, speeding up of the relaxation at a well defined field $H=1025\ Oe$, and an increase of the magnetization each time this field is crossed during a succession of minor hysteresis loops.  The observation of QTM in such a compound is compelling since here we are concerned with a large number of spins in a 3D magnetically ordered compound. More interesting is the existence of the resonance field, which supposes a crossing of discrete energy levels. In this system, the origin of these levels is not obvious since there is no well-defined entity like the macrospin as is the case for SMMs and we thus expect the process of relaxation to be very different.
 
We propose a mechanism to describe this behavior using a simple model of 0D domain-wall nucleation on an ensemble of weakly coupled ferromagnetic Ising chains. The axial anisotropy $D$ is much larger than the rhombic term  $E$ and as a consequence the spins in the domain wall will remain perpendicular to the $b$ axis, but spiral about this axis. For a wall length of $n$ spins, we assume that each spin is rotated by $\pi/n$ rad with respect to their neighbors.  Ignoring the weak interchain exchange, $n$ is found by minimizing the wall energy:
\vspace{-5mm} 
$$E_{wall}(n)=2JS_{eff}^2\ n \left( 1- \cos \left( \pi/n \right) \right) +   E S^2 \sum_{n'=1}^{n}{\sin^2 \left( n'\pi/n \right) } $$
\vspace{-5mm}

\noindent where the first term is the difference in exchange energy and the last term is the increase in anisotropy energy.  For a domain wall at the edge of a chain, using $J=14 \ K$ and $ES^2 =0.95 \ K$ gives a wall length of $n=8$ spins and wall energy of about $E_{wall}/k_B =8 \ K$. For the nucleation of a wall within a chain, all the values are doubled. In either case, the number of spins in the length of the wall is small enough so that quantum effects may be observed. Note also that the chirality of the domain walls could play an important role due to interference from the two winding paths. 
 
As with the case for SMMs, the resonance effect may be due to level crossings as we outline in the following. At $H=0$, the ground state is antiferromagnetic  and the first excited state, \textit{i.e.}, the antiferromagnetic state with one (minimum energy) domain wall, will be $E_{wall}$ above the ground state. When measuring the first magnetization curve $M_{ZFC}=0$, as H is increased these two levels will cross when $E_{wall} - ng \mu_B S H \ =0$ (the Zeeman term comes from the uncompensated spins of the domain wall). Using our experimental value for the resonance $H=1025\ Oe$, with $S_{eff}=1/2$ and $g=7.5 $ gives the ratio $n/ E_{wall} = 3.8\ k_B/K$, larger by a factor of nearly 4 than the values  calculated above. On the other hand if we assume that the activation energy obtained from Fig.~\ref{fig3} corresponds to an experimental determination of the nucleation energy, $E'/k_B =E_{wall}/k_B=2.2\ K$, then the level crossing at $H=1025\ Oe$ implies $n=8 \to 9$ in intriguing agreement with our estimate from above. At the resonance, each nucleation of a domain wall on a given chain is an independent event, governed by some characteristic relaxation time. However,  as soon as the wall is created, it will rapidly sweep along the chain reversing spins from one end of the sample to the other, (or at least until pinned by a defect or impurity), thereby greatly amplifying the relaxation for each nucleation event. 
 
Similar arguments show that for major hysteresis loops starting from the ferromagnetic down state (\textit{i.e.}, all $N$ spins pointing down and $M=-M_{sat}$) and upon increasing the field, a level crossing with  the  "ferromagnetic down state + one domain wall" will occur at  $ Ng \mu_B S H =  E_{wall} + (N- n)g \mu_B S H $ (again, ignoring the weak interchain exchange). This results in the same ratio for  $n/ E_{wall} = 3.8\ k_B/K$ as above. Notwithstanding, at the resonance field, long relaxation measurements starting from $M=-M_{sat}$ indicated that at least two characteristic relaxation times are present, implying  that the relaxation from the saturated state is a two-step process. Thus starting from $M=-M_{sat}$ the magnetization first relaxes rapidly  toward $M=0$ with a characteristic  time $\tau = 670$ s, but slows down after crossing the  $M=0$ axis, and continues to relaxes away from the antiferromagnetic state toward $M=+M_{sat}$, but with a much longer characteristic time $\tau =35000$ s.

In conclusion, we have shown that at low temperature, CoTAC exhibits spectacular properties reminiscent of SMMs. Our crude model of domain wall nucleation can explain many of the observations, but certainly needs refinement. We  cannot rule out other explanations and thus we hope our initial study will spur new interest both theoretical and experimental in this fascinating system.

We gratefully acknowledge J.-J. Pr\'{e}jean, D. Lee and W. Wernsdorfer for fruitful discussions.

\end{document}